\documentclass[aps,prd,twocolumn,showpacs,groupedaddress,floatfix,letter,nofootinbib]{revtex4-1}

\usepackage{graphicx}
\usepackage[symbol*]{footmisc}
\usepackage{array}
\usepackage{amsmath}
\usepackage{amssymb}
\usepackage{amsfonts}
\usepackage{verbatim} 

\DeclareGraphicsExtensions{.eps, .eps, .jpg, .png, .pdf}

\def\bk{{\bf k}}

\def\la{~\mbox{\raisebox{-.6ex}{$\stackrel{<}{\sim}$}}~}
\def\ga{~\mbox{\raisebox{-.6ex}{$\stackrel{>}{\sim}$}}~}
\def\N{{\cal N}}
\def\x{{\bf x}}
\def\k{{\bf k}}
\def\q{{\bf q}}
\def\p{{\bf p}}
\begin{document}

\title{Particle production during inflation and gravitational waves detectable by ground-based interferometers}

\author{Jessica L. Cook}
\email[]{jlcook@physics.umass.edu}
\author{Lorenzo Sorbo}
\email[]{sorbo@physics.umass.edu}

\affiliation{Department of Physics, University of Massachusetts, Amherst, MA 01003}


\begin{abstract}
Inflation typically predicts a quasi scale-invariant spectrum of gravitational waves. In models of slow-roll inflation, the amplitude of such a background is  too small to allow direct detection without a dedicated space-based experiment such as the proposed BBO or DECIGO. In this paper we note that particle production during inflation can generate a feature in the spectrum of primordial gravitational waves. We discuss the possibility that such a feature might be detected by ground-based laser interferometers such as Advanced LIGO and Advanced Virgo, which will become operational in the next few years. We also discuss the prospects of detection by a space interferometer like LISA.  We first study gravitational waves induced by nonperturbative, explosive particle production during inflation, finding that explosive production of scalar or vector quanta does not generate a significant bump in the primordial tensor spectrum. On the other hand, chiral gravitational waves produced by electromagnetic fields amplified by an axion-like inflaton could be detectable by Advanced LIGO.

\end{abstract}


\maketitle

\section{Introduction.} Once we consider perturbations on the top of a homogeneous and isotropic Friedmann-Robsertson-Walker Universe, inflation generically predicts quasi-scale-invariant spectra of scalar and tensor perturbations.

The scalar perturbations have been detected, and all their properties appear to wonderfully agree with the predictions of the simplest models of inflation. The tensor modes, however, have not yet been detected, and we put our best hopes to find them in the study of their effect on the polarization of the Cosmic Microwave Background photons. A {\em direct} detection of the tensor modes from inflation, on the other hand, is not expected before dedicated space-based interferometers such as the proposed BBO or DECIGO~\cite{decigo}, are launched in the next few decades. Indeed, gravitational interactions are so weak that we have not yet detected gravitational waves of {\em any} origin.

Experiments searching for gravitational waves of astrophysical origin, such as LIGO~\cite{ligo}, GEO600~\cite{geo600}, Virgo~\cite{virgo}, and TAMA300~\cite{tama300}, have been taking data for several years. LIGO and Virgo will see their sensitivity improved by a factor of $\sim 10$ in the next few years and might detect the first gravitational wave as early as 2015. They will be sensitive to a stochastic background of gravitational waves whose logarithmic contribution to the critical density, $\Omega_{GW}\,h^2$, is of the order of $10^{-9}$ at a frequency of $\sim 100$~Hz. LGCT~\cite{lcgt} will have a comparable sensitivity at similar frequencies. A space-based experiment like LISA would be able to reach $\Omega_{GW}\,h^2\simeq 10^{-11}$ at $10^{-3}$~Hz. The proposed Einstein Telescope~\cite{einstein} would have similar sensitivity while working at LIGO frequencies. Non-detection of tensor modes at CMB scales strongly constrains a scale invariant background of tensors of inflationary origin to $\Omega_{GW}\,h^2\la 10^{-14}$. Since the spectrum of primordial tensor modes is generically flat or slightly red, none of these detectors is usually expected to be able to detect tensor modes produced during inflation~\cite{Smith:2005mm}.

In this paper we note that production of particles during inflation generates a feature in the tensor spectrum which could be detectable by gravitational interferometers in the (relatively) near future, without conflicting with CMB constraints.

Nonperturbative production of particles during inflation, first studied in~\cite{Chung:1999ve}, is possible because the rolling inflaton provides a time-dependent background that carries the energy necessary for the production of light species. The simplest and most studied example of such a system is given by a scalar field $\chi$ whose mass  depends on the inflaton $\phi$. If $\chi$ becomes effectively massless as the inflaton rolls down its potential, then it becomes energetically cheap to produce its quanta. In this case, particle production happens at a precise moment during inflation determined by the time when the total mass of $\chi$ crosses zero. A second possibility is that the inflaton $\phi$ couples to a derivative of some field such as a gauge field~\cite{Garretson:1992vt}. In this case, the field can stay massless as the inflaton rolls down its potential, and particle production can happen steadily during inflation.

Particles produced through these mechanisms carry energy-momentum tensor $T_{\mu\nu}$, which perturbs the FRW metric into:
\begin{equation}
g_{\mu \nu} = a(\tau)^2(-d \tau^2 + (\delta_{ij} + h_{ij})\,dx^i dx^j) \,,
\end{equation} 
where we use conformal time \footnote{Throughout the paper we will denote by a prime a derivative with respect to the conformal time $\tau$, and by a overdot a derivative with respect to the cosmological time $t$: ${}'\equiv d/d\tau$,~$\dot{}\equiv d/dt$.} $\tau$ and a transverse, traceless decomposition, ignoring perturbations which do not source gravitational waves. 
The equation of motion of the perturbations is:
\begin{equation}\label{tenseq}
h_{ij}''+2\,\frac{a'}{a}\,h_{ij}'-\Delta\,h_{ij}=\frac{2}{M_P^2}\,\Pi_{ij}{}^{ab}\,T_{ab}\,\,, 
\end{equation}
where $\Pi_{ij}{}^{lm}=\Pi^i_l\,\Pi^j_m-\frac{1}{2}\Pi_{ij}\,\Pi^{lm}$ is the transverse, traceless projector, and $\Pi_{ij}=\delta_{ij}-\partial_i\,\partial_j/\Delta$. For any given expression of $T_{ab}(\x,\,\tau)$, one can then solve formally eq.~(\ref{tenseq}) as
\begin{equation}\label{tenssol}
h_{ij}(\k,\,\tau)=\frac{2}{M_P^2}\int d\tau' G_k(\tau,\,\tau')\,\Pi_{ij}{}^{ab}(\k)\,T_{ab}(\k,\,\tau')\,\,,
\end{equation}
where $G_k(\tau,\,\tau')$ is the retarded propagator solving the homogeneous transform of eq.~(\ref{tenseq}).  In this paper we will assume a de Sitter background $a(\tau)=-\left(H\,\tau\right)^{-1}$, so that the retarded propagator reads
\begin{align}
G_k(\tau,\tau')&=\frac{1}{k^3\,\tau'{}^2}\Big[\left(1+k^2\,\tau\,\tau'\right)\sin k\left(\tau-\tau'\right) \,+\nonumber\\
&+ \, k\left(\tau'-\tau\right) \,\cos k\left(\tau-\tau'\right)\Big]\,\Theta\left(\tau-\tau'\right)\,.
\end{align}

We will examine several ways of generating a nonvanishing $T_{ab}$ in eq.~(\ref{tenssol}), some of which induce a significant feature in the spectrum of tensor modes. If particle production happens explosively at a precise time during inflation (as in~\cite{Chung:1999ve}), the spectrum of gravitational waves will show a feature at the scale corresponding to the time of particle production. If particle production happens continuously, on the other hand, then the spectrum of induced tensor modes will be smoother. It is worth noting that, since the source of the  gravitational waves is quadratic in a gaussian field, $h_{ij}$ is expected to have a maximal three point function $\langle hhh\rangle\simeq \langle hh\rangle^{3/2}$~\cite{Adshead:2009bz,Maldacena:2011nz,Soda:2011am,Shiraishi:2011st}. The direct detectability of tensor modes produced during inflation was also considered by~\cite{Boyle:2007zx}, where it was shown that, if the post-inflationary Universe is dominated by a fluid stiffer than radiation, the primordial tensors amplitude can be significantly enhanced. More recently,~\cite{Chialva:2010jt} has studied the detectability of gravitational waves produced by phase transitions during inflation. 

In section 2 we show that the explosive production of quanta of a scalar field $\chi$ can only generate a tiny correction to the background, quasi-scale invariant spectrum of tensor modes. A scenario -- ``trapped inflation'' -- where explosive production of particles occurs several times per efolding of inflation has been considered in~\cite{Green:2009ds}. In this scenario, particle production slows down the rolling of the inflaton so that inflation can occur even on a (relatively) steep potential.  Our analysis will allow, in section 2 C, the evaluation of the amplitude of the tensors induced by the trapping fields.

In section 3 we study a mechanism analogous to that of section 2, where the scalar field $\chi$ is replaced by a vector field $A_\mu$. The amplitude of the gravitational waves produced in this case is equal to that found in section 2, multiplied by a factor two that accounts for the two helicity modes of the vector field.

Finally, in section 4 we discuss the case of tensor modes produced through a gauge field coupled  to an axion-like inflaton (as discussed in~\cite{Barnaby:2010vf,Barnaby:2011vw} and, in greater detail, in~\cite{Sorbo:2011rz}). In this case the amplitude of gravitational waves can increase dramatically at smaller scales so that the system can obey the WMAP constraint on primordial tensors at CMB scales and still lead to detectable tensors at scales probed by ground-based laser intereferometers.
\section{Gravitational waves from sudden production of scalars during inflation.}

Several systems can lead to the production of particles during inflation. The one that has received the most attention is described by the lagrangian
\begin{equation}
{\cal L}_{\phi\chi}=-\frac{1}{2}\partial_\mu\phi\,\partial^\mu\phi-V(\phi)-\frac{1}{2}\partial_\mu\chi\,\partial^\mu\chi-\frac{g^2}{2}\,\left(\phi-\phi_0\right)^2\,\chi^2\,,
\end{equation}
where $V(\phi)$ is the potential supporting inflation and where we neglect for simplicity the self-interactions of the field $\chi$. If the inflaton $\phi(\tau)$, while slowly rolling down $V(\phi)$, crosses the value $\phi_0$, then the field $\chi$ becomes momentarily massless, and its quanta can be copiously produced. The analysis of~\cite{Kofman:1997yn} shows that the occupation number of $\chi$, shortly after $\phi$ crossed $\phi_0$, is given by $n_\chi(k)=\exp\left\{-\pi \frac{k^2}{g\,|\dot\phi_0|}\right\}$, where $\dot\phi_0=d\phi/dt$ at the time $\phi$ crosses $\phi_0$.

In this section we compute the number of gravitons produced by these quanta of $\chi$. The spatial part of the stress-energy tensor for the field $\chi$ is given by $T_{ab}=\partial_a\chi\,\partial_b\chi+\delta_{ab}(\dots)$, where the part proportional to $\delta_{ab}$ is projected away by $\Pi_{ij}{}^{ab}$. We promote the field $\chi(\x,\,\tau)$ to an operator $\hat{\chi}(\x,\,\tau)$, which we Fourier transform, factoring one power of the scale factor $a(\tau)$ for canonical normalization
\begin{equation}
\hat\chi(\x,\,\tau)=\frac{1}{a(\tau)}\int\frac{d^3\k}{(2\pi)^{3/2}}\,e^{i\k\x}\,\hat\chi(\k,\,\tau)\,.
\end{equation}
Plugging this decomposition into eq.~(\ref{tenssol}) gives the tensor spectrum
\begin{align}\label{hhlong}
\langle &h_{ij}(\k,\,\tau)  h_{ij}(\k',\,\tau)\rangle=\frac{1}{2\,\pi^3\,M_P^4}\,\int\frac{d\tau'}{a(\tau')^2}G_k(\tau,\,\tau') \, \times \nonumber\\
&\times \,\int\frac{d\tau''}{a(\tau'')^2} \,G_{k'}(\tau,\,\tau'')\,\Pi_{ij}{}^{ab}(\k)\,\Pi_{ij}{}^{cd}(\k') \, \times  \nonumber\\
& \times \, \int d^3\p\,d^3\p'\,p_a  {\bf (\,k_b - \,p_b)}\,p'_c {\bf (\,k^{'}_d - \,p'_d)}  \, \times \\
& \times \,\langle\hat\chi(\p,\,\tau') \hat\chi(\k-\p,\,\tau') \hat\chi(\p',\,\tau'')\hat\chi(\k'-\p',\,\tau'')\rangle\,\nonumber.
\end{align}
The quantity $\langle \dots \rangle$ in the equation above can be reduced using Wick's theorem and ignoring the disconnected term proportional to $\delta^{(3)}({\bf k}) \delta^{(3)}({\bf k'})$. Hence we need only to evaluate the two-point correlators, being careful to avoid divergences.  

The evolution of the system occurs in three stages:

\vspace{0.2cm}

{\em (i)} At early times, significantly before $\phi$ reaches $\phi_0$, the Universe does not contain quanta of the $\chi$ field. No gravitational  waves are produced by $\chi$ during this epoch.

\vspace{0.2cm}

{\em (ii)}  As $\phi$ gets close to $\phi_0$, the effective mass of $\chi$, $m_\chi(t)\equiv g\,\left(\phi(t)-\phi_0\right)$, starts evolving nonadiabatically, $\dot{m}_\chi\ga m_\chi^2$. The duration $\Delta t_{\mathrm {nad}}$ of the epoch of nonadiabaticity is $\Delta t_{\mathrm {nad}}\simeq (g\,\dot{\phi}_0)^{-1/2}$, which must be much shorter than a Hubble time for production of quanta of $\chi$ to be efficient. During this short epoch, the quanta of $\chi$, while being produced, source gravitational waves.

\vspace{0.2cm}

{\em (iii)} After a time of the order of $\Delta t_{\mathrm {nad}}$ after $\phi$ has passed $\phi_0$, the mass $m_\chi(t)$ evolves adiabatically again. Even if they are not being produced, quanta of $\chi$ are still filling the Universe and source the tensors before diluting away in a few efoldings.

\vspace{0.2cm}

In the following subsection we will study the gravitational waves produced during the epoch {\em (iii)}, while in subsection 2 B we will discuss those produced during the nonadiabatic period {\em (ii)}. As we will see, the tensors produced during these two epochs have comparable amplitude.
\subsection{Tensor production during the adiabatic epoch.}
The main quantity we have to evaluate is $\langle \hat\chi(\p,\,\tau')\hat\chi(\q,\,\tau'')\rangle$. We decompose $\hat\chi(\k)$ into creation/annihilation operators as $\hat\chi(\k,\,\tau)=\chi(\k,\,\tau)\,\hat{a}_\k+\chi^*(-\k,\,\tau)\,\hat{a}^\dagger_{-\k}\,,$ where the function $\chi$ must obey the equation 
\begin{equation}\label{eqchik}
\chi''(\k,\,\tau)+\omega(\k,\,\tau)^2\,\chi(\k,\,\tau)=0\,
\end{equation}
with
\begin{equation}\label{defomega}
\omega(\k,\,\tau)^2\equiv k^2+g^2\,\,a(\tau)^2\left(\phi(\tau)-\phi_0\right)^2 -\,\frac{a''(\tau)}{a(\tau)}\, ,
\end{equation}
and $\hat{a}_\k$ annihilates the vacuum during period {\em (i)}.
We define the Bogolyubov coefficients $\alpha(\k,\,\tau)$ and $\beta(\k,\,\tau)$ via
\begin{align}
\chi(\k,\,\tau)=&\frac{1}{\sqrt{2\,\omega}}\left(e^{-i\int^\tau \omega}\,\alpha(\k,\,\tau)+e^{i\int^\tau \omega}\,\beta(\k,\,\tau)\right)\nonumber\\
\chi'(\k,\,\tau)=& i\,\sqrt{\frac{\omega}{2}}\Big(-e^{-i\int^\tau \omega }\,\alpha(\k,\,\tau)+ e^{i\int^\tau \omega}\,\beta(\k,\,\tau)\Big)
\end{align}
so that, in the adiabatic limit $\omega'\ll\omega^2$, $\alpha$ and $\beta$ are constants. This way, we can rewrite
\begin{equation}\label{chihat}
\hat\chi(\k,\,\tau)=\frac{e^{-i\int^\tau \omega\, d\tilde\tau}}{\sqrt{2\,\omega}}\,\hat{b}_\k(\tau)+\frac{e^{i\int^\tau \omega\, d\tilde\tau}}{\sqrt{2\,\omega}}\,\hat{b}^\dagger_{-\k}(\tau)\,,
\end{equation}
where we have defined the new annihilation operator
\begin{equation}\label{decohat}
\hat{b}_\k(\tau)=\alpha(\k,\,\tau)\,\hat{a}_\k+\beta^*(-\k,\,\tau)\,\hat{a}_{-\k}^\dagger\,,
\end{equation}
that annihilates the vacuum during period {\em (iii)}.

In order to renormalize the theory, we impose that the operator $\hat\chi(\p,\,\tau')\hat\chi(\q,\,\tau'')$ within $\langle \dots \rangle$ is normal ordered. However, {\em we require normal ordering in terms of the $\hat{b}_\bk$ operators} while using the vacuum state defined by the $\hat{a}_\bk$ operators. This way we calculate the number of quanta of our initial particle definition existing at the end. Using the decomposition~(\ref{decohat}) and the commutation relation $[\hat{a}_\q,\,\hat{a}^\dagger_\p]=\delta^{(3)}(\p-\q)$, we obtain
%
\begin{widetext}
\begin{align}\label{twoptchi}
\langle\hat\chi(\p,\,\tau')\hat\chi(\q,\,\tau'')\rangle=\frac{\delta^{(3)}(\p+\q)}{2\,\sqrt{\omega_\p(\tau')\,\omega_\p(\tau'')}} &\left[\left(e^{i\int_{\tau'}^{\tau''}\omega_\p } \,\beta^*(-\p,\,\tau')\beta(-\p,\,\tau'')+ {\rm h.c.}\right)\right.\\
& +\left.\left( e^{-i \int^{\tau'} \omega_\p -i \int^{\tau''} \omega_\p } \alpha({\bf p}, \tau') \beta^*({\bf p}, \tau'') +\left (\tau'\leftrightarrow \tau'', {\mathrm {h.c.}}\right) \right)  \right]\,.\nonumber
\end{align}
Using the expression above and Wick's theorem, eq.~(\ref{hhlong}) can be written as
\begin{eqnarray}\label{twopt}
&\langle h_{ij}(\k,\,\tau)\,h_{ij}(\k',\,\tau)\rangle=\frac{\delta^{(3)}(\k+\k')}{8\,\pi^3\,M_P^4}\int d^3\p\left(p^2-\frac{\left(\p\, \cdot \k\right)^2}{k^2}\right)^2\,\int\frac{d\tau'}{a(\tau')^2}\frac{G_k(\tau,\,\tau')}{\sqrt{\omega_\p(\tau')\,\omega_{\k-\p}(\tau')}}\,\int\frac{d\tau''}{a(\tau'')^2}\frac{G_k(\tau,\,\tau'')}{\sqrt{\omega_\p(\tau'')\,\omega_{\k-\p}(\tau'')}}  \nonumber\\
&\times[e^{i\int_{\tau'}^{\tau''}\omega_\p}\beta^*(|\p|,\,\tau')\,\beta(|\p|,\,\tau'') + e^{-i \int^{\tau'} \omega_\p} e^{-i \int^{\tau''} \omega_\p} \alpha(|{\bf p}|, \tau') \beta^*(|{\bf p}|, \tau'') +  \left (\tau'\leftrightarrow \tau'', {\mathrm {h.c.}}\right)  \Big]\\
&\times \,\Big[e^{i\int_{\tau'}^{\tau''}\omega_{\k-\p} }\beta^*(|\k-\p|,\,\tau')\,\beta(|\k-\p|,\,\tau'')+ e^{i \int^{\tau'} \omega_{\k-\p}} e^{-i \int^{\tau''} \omega_{\k-\p}} \alpha(|\k-\p|, \tau') \beta^*(|\k-\p|, \tau'') +  \left (\tau'\leftrightarrow \tau'', {\mathrm {h.c.}}\right)  \Big] \,\nonumber\,.
\end{eqnarray}
\end{widetext}
%
When multiplied out, some terms in the above equation are rapidly oscillating and we neglect them as they give subdominant contribution to the integrals. At this point we need to evaluate the Bogolyubov coefficients $\alpha$ and $\beta$.
\subsubsection{Evaluating $\beta$.}
The function $\chi$ obeys eq.~(\ref{eqchik}), and quanta of $\chi$ are produced during the short epoch of nonadiabatic evolution of $\omega(\k,\,\tau)$, during which we neglect the expansion of the Universe. In a de Sitter background and in the slow-roll approximation, $\phi$ evolves approximately linearly in physical time $\phi(t) = \phi_0 + \dot{\phi}_0 \, t$ or as $\phi(\tau) = \phi_0 - \frac{\dot{\phi}_0}{H} \log(\frac{\tau}{\tau_0})$ in conformal time. $t$ and $\tau$ are defined such that $\phi(t=0) = \phi(\tau= \tau_0) = \phi_0$. The duration $\Delta t_{\mathrm {nad}}$ of the nonadiabatic epoch is determined by the condition $\dot{m}_\chi\ga m^2_\chi$, yielding $\Delta t_{\mathrm {nad}}\sim 1/\sqrt{g|\dot\phi_0|}$. It is consistent to neglect the expansion of the Universe if $\Delta t_{\mathrm {nad}}\ll 1/H$ so that the validity of our analysis requires $g\gg H^2/|\dot\phi_0|$.

Under these conditions the equation for $\chi$ during the nonadiabatic epoch reduces to
\begin{equation}\label{eqchinad}
\ddot\chi+\left(k^2\,H^2\,\tau_0^2+g^2\,\dot\phi_0^2\,t^2\right)\,\chi=0\,\,,
\end{equation}
to which we can apply the analysis of~\cite{Kofman:1997yn}, obtaining, up to an irrelevant phase, the Bogolyubov coefficients
\begin{eqnarray}
&&\alpha(\tau>\tau_0,k)=\sqrt{1+e^{-\pi\,\kappa^2}}\,e^{i\,\alpha_\kappa}\,,\nonumber\\
&&\beta(\tau>\tau_0, k)=e^{-\frac{\pi}{2}\,\kappa^2}\,,
\end{eqnarray}
where we have defined  $\kappa\equiv k\,H\,\tau_0/\sqrt{g\,\dot{\phi}_0}$ and $\alpha_\kappa={\mathrm {Arg}}\left[\Gamma\left(\frac{1+i\,\kappa^2}{2}\right)\right]+\frac{\kappa^2}{2}\,\left(1-\log\frac{\kappa^2}{2}\right)$.

\subsubsection{The two point function.}
After the phase of nonadiabatic evolution of $m_\chi$, the Universe contains $\sim\int d^3\k\,|\beta|^2/a(\tau)^3$ quanta of $\chi$ per unit volume. We are now in the position of computing the spectrum of gravitational waves generated by such a population, which is given by the integral~(\ref{twopt}), where $\omega$ is defined in~(\ref{defomega}). The expression of $\omega$ can be drastically simplified by observing that, at the end of the nonadiabatic period, $g\,\left(\phi-\phi_0\right)\sim g\,\dot\phi_0\,\Delta t_{\mathrm {nad}}\sim(g\,\dot\phi_0)^{1/2}\gg H$. As a consequence, the second term in~(\ref{defomega}) is much larger than the third one. Moreover, the exponential suppression in $\beta$ means only momenta $k\la (g\,\dot\phi_0)^{1/2}/(H\,\tau_0)$ contribute significantly to the integral. Noting also that, following the non-adiabatic period, $|\tau| < |\tau_0|$, the $k^2$ in ~(\ref{defomega}) is negligible as well.

Therefore, during the entire phase {\em (iii)} of adiabatic evolution of the system, we can approximate $\omega\simeq |g\,\left(\phi(\tau)-\phi_0\right)/(H\,\tau)|$, where $\phi(\tau)-\phi_0\simeq -(\dot\phi_0/H)\,\log(\tau/\tau_0)$. Using these estimates, we can write the integral~(\ref{twopt}) in the limit $\tau\rightarrow 0$ (i.e., at the end of inflation, long after $\tau_0$) as
\begin{widetext}
\begin{align}\label{twopt1}
&\langle h_{ij}(\k)\,h_{ij}(\k')\rangle=  \frac{\delta^{(3)}({\bf k+k'})}{4 \,\pi^3 \,k^6}  \,\frac{ H^{8}}{g^2\,\dot{\phi}^2_0\,M_P^4}   \int_{- \infty}^{\infty} d^3 {\bf p} \,  \left(p^2  - \frac{({\bf p \cdot k})^2}{k^2}\right)^2 \times\\
\times \Big\{ \left|\beta\left({\bf p}\right)\right|^2\,\left|\beta\left({\bf k}-{\bf p}\right)\right|^2+&{\mathrm {Re}}\left[\alpha({\bf p})\,\alpha^*({\bf k}-{\bf p})\,\beta^*({\bf p})\,\beta({\bf k}-{\bf p})\right]\Big\}\,\left[\int_{\tau_0+\Delta\tau_{\mathrm {nad}}}^{0} \,d\tau\, \tau\,\frac{ \sin(k\,\tau)-k\,\tau\,(k\,\tau)}{ \ln({\tau_0}/{\tau})  } \right]^2\nonumber ,
\end{align}
\end{widetext}
where $\Delta\tau_{\mathrm {nad}}/|\tau_0|=H\,\Delta t_{\mathrm {nad}}\simeq H\,(g\dot\phi_0)^{-1/2}\ll 1$. As we will see in eq.~(\ref{fepsilon}) below, the result depends only logarithmically on $\Delta t_{\mathrm {nad}}$ so that ignorance of its exact value does not affect significantly the results. Next, we recognize that at large values of $k$  the two point function is suppressed by the factor of $(\sin k \tau - k \tau \, \cos k \tau)/ k^3$ coming from the Green's functions. Since $H^2/ g \dot{\phi} \ll 1$, $p$ is only suppressed after $p \tau_0 > g \dot{\phi}/ H^2 \gg 1$. Therefore the integrand gets its main contribution from the region $p \gg k$. Using these approximations, the above equation is simplified to
\begin{align}\label{eq18}
& \langle h_{ij}(\k)\,h_{ij}(\k')\rangle= \frac{\delta^{(3)}({\bf k+k'})}{4 \,\pi^3 \,k^6}  \,\frac{ H^{8}}{g^2\,\dot{\phi}^2_0\,M_P^4}\times  \nonumber  \\
& \int d^3 {\bf p}   \left(p^2  - \frac{({\bf p \cdot k})^2}{k^2}\right)^2 \left( 2\,e^{-\frac{\pi  p^2 H^2 \tau_0^2}{g\dot{\phi}}} +   e^{-\frac{2 \pi p^2 H^2 \tau_0^2}{ g\dot{\phi}}} \right)\times\nonumber\\
& \left(\int_{\tau_0+\Delta\tau_{\mathrm {nad}}}^{0} \,d\tau\, \tau\,\frac{ \sin(k\,\tau)-k\,\tau\,(k\,\tau)}{ \ln({\tau_0}/{\tau})  } \right)^2.
\end{align}
After computing the integral in $d^3\p$, we are left with the simple expression
\begin{align}\label{twoptsemi}
\langle h_{ij}(\k)\,h_{ij}(\k')\rangle&=\frac{\delta^{(3)}(\k+\k')}{2\,\pi^5\,k^6\,|\tau_0|^3}\,\frac{H^4}{M_P^4}\, \left(1+\frac{1}{4 \sqrt{2}} \right)\times\nonumber\\
&\times\left(\frac{g\,\dot\phi_0}{H^2}\right)^{3/2}\,F_{|\Delta\tau_{\mathrm {nad}}/\tau_0|}\left(k|\tau_0|\right)\, ,
\end{align}
where 
\begin{align}\label{fepsilon}
F_\epsilon(y) &\equiv \left|\int_0^{1-\epsilon} x\,\frac{\left(\sin xy-xy\,\cos xy\right)}{\log x}\,dx\right|^2\underset{\epsilon \to 0}{\simeq} \nonumber\\
& \simeq \left[\left(y\cos y-\sin y\right)\,\log\epsilon\right]^2\,.
\end{align}

The two point function~(\ref{twoptsemi}) should be added to the standard, quasi scale invariant contribution from inflationary gravitational waves so that the resulting power spectrum reads
\begin{align}
{\cal P}^t(k)&\simeq\frac{2\,H^2}{\pi^2\,M_P^2}\,\left[1+4.8\times 10^{-4}\,\frac{\left(k\tau_0\cos k\tau_0-\sin k\tau_0\right)^2}{\left|k\,\tau_0\right|^3}\,\right. \times \nonumber\\
&\left. \times\,\frac{H^2}{M_P^2}\, \left(\frac{g\,\dot\phi_0}{H^2}\right)^{3/2}\,\log^2\left(\frac{\sqrt{g\,\dot\phi_0}}{H}\right)\right]\, .
\end{align}

We thus see that the effect of the creation of quanta of $\chi$ is to superimpose a scale dependent contribution to the scale invariant spectrum of tensors generated by inflation. 

We next observe that $\dot\phi_0=\sqrt{2\,\epsilon}\,H\,M_P$, where $\epsilon\ll 1$ is the slow-roll parameter. Supplying reasonable values for $H$, $M_P$, and $\epsilon$ allows us to find that the $|k\tau_0|$-dependent part of the spectrum is maximized at $|k\tau_0|\simeq 2$, where the component from particle production evaluates to $\sim 1.8\times 10^{-4}\,\frac{H^2}{M_P^2}\,(\frac{g\dot\phi_0}{H^2})^{3/2}\,\log^2(\frac{\sqrt{g\dot\phi_0}}{H})$. Using the same approximation, the $\log^2$ term gives at most a factor $\sim 10^2$; therefore, the correction to the standard result is at most of the order $10^{-2}\,H^{1/2}/M_P^{1/2}$, which is several orders of magnitude smaller than unity.

We thus conclude that the presence of a gas of adiabatically evolving scalar particles produced nonperturbatively during inflation generates a tiny correction to the spectrum of primordial tensors. This result agrees with~\cite{Dufaux:2007pt}, where it was shown that in the Minkowsky limit, $H\to 0$, a gas of adiabatically evolving scalars does not generate any gravitational waves. In our case, since we are on an expanding background, gravitational waves are produced, but the effect is still small and unobservable.

Let us next estimate the amount of gravitational waves produced during the period of nonadiabatic evolution of $m_\chi(\tau)$.
\subsection{Tensor production during the nonadiabatic epoch.}
The period of nonadiabatic evolution of $m_\chi(\tau)$ lasts much less than a Hubble time.  We will use again our physical time variable $t=H^{-1}\,\log(\tau_0/\tau)$. Since we are now looking at the period $|H\,t|\ll 1$, we can approximate the change of variable as $\tau\simeq \tau_0\left(1-H\,t\right)$. This implies that we can replace $\tau'$ and $\tau''$ by $\tau_0$ in the integrands of eq.~(\ref{hhlong}). During this short time, the field $\chi$ will obey eq.~(\ref{eqchinad}).

During the periods of adiabatic evolution of $m_\chi$, the concept of a particle of $\chi$ is well defined, and the use of the Bogolyubov coefficients gives an appropriate way of computing the spectrum of gravitons produced by the gas of quanta of $\chi$. During the short epoch of nonadiabatic evolution of $m_\chi$, however, it is more convenient to switch to a different prescription. Following e.g.~\cite{Barnaby:2009mc}, we  set
\begin{align}\label{newpre}
\langle\hat\chi(\p,\,t')\hat\chi(\q,\,t'')\rangle&=\delta^{(3)}(\p+\q)\,\Big[\chi(\p,\,t')\chi^*(\p,\,t'') - \nonumber\\
& -\tilde\chi(\p,\,t')\tilde\chi^*(\p,\,t'')\Big]\,
\end{align}
where $\chi(\p,\,t)$ is the solution, with appropriate boundary conditions, to eq.~(\ref{eqchinad}), and $\tilde\chi(\p,\,t)$ is the solution to the same equation in the adiabatic approximation:
\begin{eqnarray}\label{defchinad}
&&\chi(\p,\,t)=\frac{\sqrt{H\,\tau_0}}{(g\,\dot\phi_0)^{1/4}}\,e^{-i\,\pi/8}\,e^{-\pi\,\bar{p}^2/8}\,D_{\frac{-1+i\bar{p}^2}{2}}\left[(-1+i)\eta\right]\,\,,\nonumber\\
&&\tilde\chi(\p,t)=\frac{\sqrt{H\tau_0}}{(g \dot\phi_0)^{1/4}} \frac{e^{i\left[\frac{\eta}{ 2}\sqrt{\eta^2+\bar{p}^2}+\bar{p}^2/2 \log (\eta/\bar{p}+\sqrt{1+\eta^2/\bar{p}^2})\right]}}{\sqrt{2\,\sqrt{\eta^2+\bar{p}^2}}}\,,\nonumber\\
\end{eqnarray}
where we have defined a dimensionless time $\eta=(g\,\dot\phi_0)^{1/2}\,t$ and a  dimensionless momentum $\bar{p}=p\,H\,\tau_0/(g\,\dot\phi_0)^{1/2}$ and $D_a(z)$ is the parabolic cylinder function. The term in $\tilde\chi(\p,\,t)$ in eq.~(\ref{newpre}) takes care of the UV-divergent terms which would otherwise appear in the tensor spectrum.

Working forward from eq.~(\ref{newpre}), the expression for the contribution of the nonadiabatic epoch to the two point correlator of the graviton simplifies to
%
\begin{widetext}
\begin{align}\label{hhff}
\langle h_{ij}&(\k,\,\tau)\,h_{ij}(\k',\,\tau)\rangle=\frac{\delta^{(3)}(\k+\k')}{2\,\pi^3\,M_P^4}\frac{H^4}{k^6}\,\Big[\sin k\,\tau_0-k\,\tau_0\,\cos k\,\tau_0\Big]^2 \int d^3\p\,\left(p^2-\frac{(\p \cdot \k)^2}{k^2}\right)^2 \,(H\,\tau_0)^2 \\
&\times\, \int dt'dt''\left[\chi(\p,\,t')\chi^*(\p,\,t'')-\tilde\chi(\p,\,t')\tilde{\chi}^*(\p,\,t'')\right]\, \left[\chi(\k-\p,\,t')\chi^*(\k-\p,\,t'')- \tilde\chi(\k-\p,\,t')\tilde{\chi}^*(\k-\p,\,t'')\right]\,.\nonumber
\end{align}
\end{widetext}
%
We next observe that for $|k\,\tau_0| \gg 1$ the two point function of the graviton is suppressed by the coefficient $\propto k^{-4}$ in front of the integrals of eq.~(\ref{hhff}). In the unsuppressed regime $|k\,\tau_0| \la 1$, the quantity $\bar{k}=k\,H\,|\tau_0|/(g\,\dot\phi_0)^{1/2}\ll 1$, so that we can set $\k=0$ in the second line of eq.~(\ref{hhff}). At this point  the angular integral can be easily computed, and by observing that both $\chi$ and $\tilde\chi$ in eq.~(\ref{defchinad}) are of the form $(H\,|\tau_0|)^{1/2}\,(g\,\dot\phi_0)^{-1/4}\times[$function of $(\bar{p},\,\eta)]$, we can write, after appropriate changes of variables,
\begin{align}\label{hhff2}
\langle h_{ij}(\k,\,\tau)&\,h_{ij}(\k',\,\tau)\rangle=\frac{16}{15\,\pi^2}\,\frac{\delta^{(3)}(\k+\k')}{k^6\,|\tau_0|^3}\frac{H^4}{M_P^4}\, \times \nonumber\\
& \times \left(\frac{g\,\dot\phi_0}{H^2}\right)^{3/2}\,\left[\sin k\,\tau_0-k\,\tau_0\,\cos k\,\tau_0\right]^2\times\nonumber\\
&\times\int \bar{p}^6\,d\bar{p}\,\int d\eta'\,d\eta''\,F(\bar{p},\,\eta',\,\eta'')^2\, ,
\end{align}
where $F(\bar{p},\,\eta',\,\eta'')$ in the second line is a dimensionless function built out of the dimensionless functions appearing in eqs.~(\ref{defchinad}), whose integral gives an ${\cal O}(1)$ result.

The result of this subsection is that the contribution to the two-point function of the graviton by the nonadiabatic epoch {\em (ii)} has the same form, modulo a logarithmic term, as the contribution from the adiabatic epoch discussed in section 2 A. We conclude that the overall effect of graviton creation by the scalars $\chi$ gives a negligible correction to the standard spectrum of tensors generated by inflation. Let us next discuss what happens if there are several events of explosive production of scalars.
\subsection{An application: tensor modes in trapped inflation.}
The analysis of sections 2 A and 2 B has shown that the spectrum of tensor modes induced by a single event of production of scalars has the form
\begin{align}
{\cal P}^t_\chi&\simeq \frac{H^4}{M_P^4}\frac{\left(k|\tau_0|\,\cos k|\tau_0|-\sin k|\tau_0|\right)^2}{(2\,\pi)^3\,k^3\,|\tau_0|^3}\, \left(\frac{g\,\dot\phi_0}{H^2}\right)^{3/2}
\end{align}
up to a coefficient of order one. Let us now suppose that there are several instances of particle production. In particular, we will assume that particle production happens so often to lead to trapped inflation~\cite{Green:2009ds}, i.e., to a slowing down of the rolling of the inflaton more significant than that due to Hubble friction\footnote{The mechanism leading to trapped inflation is similar to that responsible for warm inflation~\cite{astro-ph/9509049,arXiv:0808.1855}, although in the case of warm inflation friction is provided by a thermal bath rather than by nonperturbative particle production.}. Since the tensor modes excited by each burst of particles sum incoherently, the resulting power spectrum of the tensor will just be the sum of the individual power spectra. If the bursts happen frequently enough, summation the various contributions to ${\cal P}^t_\chi$ is equivalent to integration over $\frac{d\tau_0}{\Delta}\,\frac{d\phi}{d\tau_0}$, where $\Delta=\phi_{i+1}-\phi_i$ is the distance in field space between two sites of particle production. We therefore get,
\begin{align}
{\cal P}^t_{\mathrm {trapped}}&\simeq \int d\tau_0\frac{a(\tau_0)\,\dot{\phi}_0}{\Delta}\,\frac{H^4}{M_P^4}\frac{\left(k|\tau_0|\,\cos k|\tau_0|-\sin k|\tau_0|\right)^2}{(2\,\pi)^3\,k^3\,|\tau_0|^3}\, \nonumber\\
& \times\left(\frac{g\,\dot\phi_0}{H^2}\right)^{3/2}\, ,
\end{align}
i.e., in an order of magnitude estimate, 
\begin{equation}
{\cal P}^t_{\mathrm {trapped}}\simeq \frac{\dot{\phi}_0}{(2\,\pi)^3\,H\,\Delta}\,\frac{H^4}{M_P^4}\,\left(\frac{g\,\dot\phi_0}{H^2}\right)^{3/2}\,\,,
\end{equation}
which shows that the resulting spectrum is scale invariant. 

Let us now connect this result to the parameter space of trapped inflation. Plugging the "slow roll" equation of trapped inflation~\cite{Green:2009ds}, $(g\,\dot{\phi}_0)^{5/2}\simeq (2\,\pi)^3\,H\,\Delta\,V'$, into the equation for the tensors, we obtain 
\begin{equation}
{\cal P}^t_{\mathrm {trapped}}\simeq \frac{H\,V'}{g\,M_P^4}\,.
\end{equation}

This should be compared to the standard amplitude of gravitational waves ${\cal P}^t_{\mathrm {standard}}\simeq H^2/M_P^2$, so that ${\cal P}^t_{\mathrm {trapped}}/{\cal P}^t_{\mathrm {standard}}\simeq H\,V'/g\,V$. In order to proceed further, we choose a form of the potential. For $V(\phi)=m^2\,\phi^2/2$ (the case studied in~\cite{Green:2009ds}) ${\cal P}^t_{\mathrm {trapped}}/{\cal P}^t_{\mathrm {standard}}\simeq m/g\,M_P$. This is much smaller than unity in the phenomenologically allowed region of parameter space of~\cite{Green:2009ds}. Therefore, even in the case of several bursts of  scalars, the induced spectrum of tensors is subdominant with respect to the standard one $\sim H^2/M_P^2$.

\section{Production via vectors.}

In this section we study the production of gravitational waves induced by the explosive production of vectors. We will only focus on the gravitational waves produced during the period of adiabatic evolution subsequent to the creation of the vectors. Based on the similarities between the results of sections 2 A and 2 B, we expect that the contribution of the nonadiabatic period will at most give an order one correction to the results presented here.
  
We consider the gauge-invariant lagrangian
\begin{equation}
{\cal {L}}=-\frac{1}{4}\,F_{\mu\nu}\,F^{\mu\nu}-(D^\mu\Psi)\,(D_\mu\Psi)^*-V(|\Psi|^2)\, ,
\end{equation}
where $D_\mu=\nabla_\mu-i\,e\,A_\mu$ is the gauge-covariant derivative. We will assume that the Higgs field $\Psi$ is a function of time, but is otherwise homogeneous during inflation, Defining $\Psi\equiv\psi\,e^{i\theta}$, it is possible to show~\cite{Finelli:2000sh} that, if $\Psi$ is spatially homogeneous, it is consistent to choose the gauge $A_0(t,\,{\bf x})=\dot\theta(t)/e,\,\nabla\cdot{\bf A}=0$, with ${\bf A}$ satisfying the equation
\begin{equation}
{\bf A}''({\bf k},\,\tau)+\left(k^2+e^2\,a^2(\tau)\,\psi^2(\tau)\right){\bf A}({\bf k},\,\tau)=0\,,
\end{equation}
so when $\psi$ crosses zero, production of photons occurs precisely in the same way as described above in the case of scalars. In particular, the expression of the Bogolyubov coefficients is calculated the same way as that of section 2.A.1. The portion of spatial part of the stress-energy tensor of ${\bf A}$ we are interested in is given by $T_{ij} = A'_i\, A'_j + \epsilon_{ikl} \partial^k A^l\, \epsilon_{jmn} \partial^m A^n-m^2(\tau)\,A_i\,A_j$, where we have defined $m(\tau)\equiv e\,a(\tau)\,\psi(\tau)$. We promote the photon to a quantum field $\hat{\bf A}(\tau,\,{\bf x})$, which we decompose on a basis of helicity vectors ${\bf e}_\pm(\k)$ satisfying the conditions $\k\cdot {\bf e}_\pm(\k)=0$, ${\bf e}_\sigma(\k)\cdot{\bf e}_{\sigma'}(\k)=\delta_{\sigma,-\sigma'}$, ${\bf e}_\pm(-\k)=-{\bf e}_\mp(\k)$, $\k\times {\bf e}_\pm(\k)=\mp\, i\, k\, {\bf e}_\pm(\k)$ and ${\bf e}_\pm(\k)^*={\bf e}_\mp(\k)$:
\begin{align}\label{decoa}
{\hat {\bf A}}&=\sum_{\lambda=\pm}\int \frac{d^3\bk}{\left(2\pi \right)^{3/2}}\Big[{\bf{e}}_\lambda(\bk)\,\hat{A}_\lambda(\tau,\,\bk)\,e^{i{\bf k\cdot x}} + {\mathrm {h.c.}}\Big], 
\end{align}
with $\hat{A}_\lambda(\tau,\,\bk)={A}_\lambda(\tau,\,\bk)\,\hat{a}(\bk)$. Eq.~(\ref{tenssol}) then reads
\begin{align}\label{hijvec}
&{\hat h}_{\lambda}(\k,\,\tau)=\frac{2}{M_P^2}\,\sum_{\sigma,\,\sigma'}\int d\tau'\,\frac{G_k(\tau,\,\tau')}{a(\tau')^2}\int\frac{d^3{\q}}{(2\pi)^{3/2}}\\
&\times \Pi_\lambda^{lm}(\k)\,{\bf e}_\sigma^l(\q)\,{\bf e}_{\sigma'}^m(\k-\q)\Big[\hat{A}_\sigma'({\q},\tau')\,\hat{A}_{\sigma'}'({\k}-{\q},\tau') \nonumber\\
&+\left(\sigma\,\sigma'\,q\,|\k-\q|-m^2(\tau')\right)\,\hat{A}_\sigma({\q},\tau')\,\hat{A}_{\sigma'}({\k}-{\q},\tau')\Big]\,.\nonumber
\end{align}
where we have introduced the projector $\Pi_\lambda^{lm}(\k)\equiv {\bf e}_\lambda^l(\k)\,{\bf e}_{\lambda}^m(\k)/\sqrt{2}$ on the helicity-$\lambda$ component of the graviton.

We are now in position to compute the two-point function of the graviton $\langle \hat{h}_\lambda(\k,\,\tau)\,\hat{h}_{\lambda'}(\k',\,\tau)\rangle$. To do so, we use the Wick theorem to express it as a function of two-point correlators of the gauge field. Next, we define a function $f_2(\k,\,\tau',\,\tau'')$ such that
\begin{align}
\langle \hat{A}_\sigma(\p,\,\tau')\,\hat{A}_{\sigma'}(\q,\,\tau'')\equiv \delta^{(3)}(\p+\q)\,\delta_{\sigma\sigma'}\,f_2(\p,\,\tau',\,\tau'')\,.
\end{align}
The expression of $f_2(\p,\,\tau',\,\tau'')$ can be read from eq.~(\ref{twoptchi}). Since eq.~(\ref{hijvec}) contains time derivatives of $A_\sigma(\k,\,\tau')$, the two point-function of the graviton will contain terms such as $\partial_{\tau'}f_2(\q,\,\tau',\,\tau'')\,\partial_{\tau'}f_2(\k-\q,\,\tau',\,\tau'')$, that can be simplified by integration by parts in $\tau'$ or $\tau''$ and by using the property $\partial^2_{\tau'}f_2(\q,\,\tau',\,\tau'')=-\omega_q^2(\tau')\,f_2(\q,\,\tau',\,\tau'')$.

After several such integrations by parts we can write the two-point function of the graviton as 
%
\begin{widetext}
\begin{align}\label{hhaa}
&\hspace{1cm}\langle h_{\lambda}(\k,\,\tau)\,h_{\lambda'}(\k',\,\tau)\rangle=\frac{\delta^{(3)}(\k+\k')}{4\,\pi^3\,M_P^4}\,\delta_{\lambda\lambda'}\,\int d\tau'\,d\tau'' \int d^3\p\,f_2(\p,\,\tau',\,\tau'')\,f_2(\k-\p,\,\tau',\,\tau'') \\
&\times\Big\{\chi_1(\k,\,\p)\left[\tilde{G}''(\tau')\,\tilde{G}''(\tau'')+(\p^2+(\k-\p)^2)\left(\tilde{G}''(\tau')\,\tilde{G}(\tau'')+\tilde{G}(\tau')\,\tilde{G}''(\tau'')\right)+2\,(\p^2+(\k-\p)^2)^2\tilde{G}(\tau')\,\tilde{G}(\tau'')\right]\nonumber\\
&+p\,|\k-\p|\,\chi_2(\k,\,\p)\left[2\left(\tilde{G}''(\tau')\,\tilde{G}(\tau'')+\tilde{G}(\tau')\,\tilde{G}''(\tau'')\right)+4\,(\p^2+(\k-\p)^2)\,\tilde{G}(\tau')\,\tilde{G}(\tau'')\right]\Big\}\,,\nonumber
\end{align}
\end{widetext}
%
where we have defined
\begin{align}
&\chi_1(\k,\,\p)\equiv \sum_{\sigma\sigma'}\left|\Pi_\lambda^{ab}(\k)\,{\bf e}_\sigma^a(\p)\,{\bf e}_{\sigma'}^b(\k-\p)\right|^2,\nonumber\\
&\chi_2(\k,\,\p)\equiv \sum_{\sigma\sigma'}\,\sigma\,\sigma'\,\left|\Pi_\lambda^{ab}(\k)\,{\bf e}_\sigma^a(\p)\,{\bf e}_{\sigma'}^b(\k-\p)\right|^2\,,\nonumber\\
&\tilde{G}(\tau')\equiv \frac{G_k(\tau,\,\tau')}{a(\tau')^2}\,,\quad \tilde{G}''(\tau')\equiv \partial^2_{\tau'}\left(\frac{G_k(\tau,\,\tau')}{a(\tau')^2}\right)\,.
\end{align}
Up to this point all the expressions used are exact. Next, we employ the same approximations of section 2 A above: we impose $\omega_k(\tau')\simeq m(\tau')$ and we drop all terms contain fast-oscillating phases of the form $e^{2i\int m(\tilde{\tau})\,d\tilde{\tau}}$. In this regime 
\begin{align}
&f_2(\p,\,\tau',\,\tau'')\,f_2(\k-\p,\,\tau',\,\tau'') \simeq \frac{1}{4\,m(\tau')\,m(\tau'')}\\
&\left[|\beta({\bf p})|^2|\beta({\bf k}-{\bf p})|^2+{\mathrm {Re}}\left[\alpha({\bf p})\alpha^*({\bf k}-{\bf p})\beta^*({\bf p})\beta({\bf k}-{\bf p})\right]\right].\nonumber
\end{align}

Now all integrals appearing in eq.~(\ref{hhaa}) can be computed explicitly. In the limit $e\,\dot\psi/H^2\gg 1$ the terms in $\tilde{G}''(\tau')$ turn out to be subdominant with respect to those in $\tilde{G}(\tau')$. Also, the integral $\int d\tau'\tilde{G}(\tau')/m(\tau')$ is suppressed by powers of $k\,\tau_0$ for $k\,\tau_0\gg 1$, while the integral in $d^3\p$ is dominated by $p\,\tau_0\sim e\,\dot\psi/H^2\gg 1$, so that we can set $|\k|\ll |\p|$. The final result is 
\begin{align}
&\langle h_{\lambda}(\k)\,h_{\lambda'}(\k')\rangle=\frac{\delta^{(3)}(\k+\k')}{2\,\pi^3\,M_P^4}\,\delta_{\lambda\lambda'}\left(\int d\tau'\,\frac{\tilde{G}(\tau')}{m(\tau')}\right)^2\nonumber\\
&\times\int d^3\p\,p^4\left(\chi_1(\k,\,\p)+\chi_2(\k,\,\p)\right)\left(|\beta(\p)|^2+2\,|\beta(\p)|^4\right)\,.
\end{align}
The integral in $d\tau'$ appearing in the equation above has the same form as that appearing in eq.~(\ref{eq18}). Therefore we treat it as we did in eq.~(\ref{fepsilon}). Also, in the regime $|\k|\ll |\p|$ we are interested in, $\chi_1(\k,\,\p)+\chi_2(\k,\,\p)=\frac{1}{8}\left(1-(\p\,\k)^2/p^2\,k^2\right)^2$, so that the two-point function of the graviton, once we sum on the two helicity states $\lambda=\pm 1$ amounts to twice as much as that of the gravitons produced by the scalar sources~(\ref{twoptsemi}):
\begin{align}
&\sum_{\lambda\lambda'}\langle h_{\lambda}(\k)\,h_{\lambda'}(\k')\rangle\simeq\frac{\delta^{(3)}(\k+\k')}{k^3}\,\frac{H^4}{\pi^5\,M_P^4}\, \left(1+\frac{1}{4 \sqrt{2}} \right)\nonumber\\
&\times\left(\frac{g\,\dot\psi}{H^2}\right)^{3/2}\frac{\left(k\tau_0\cos k\tau_0-\sin k\tau_0\right)^2}{k^3\,|\tau_0|^3}\,\log^2\frac{\sqrt{g\,\dot\psi}}{H}\,.
\end{align}

We conclude that the amplitude of the gravitational waves induced by vectors equals, per helicity mode, that induced by scalars and is therefore negligible.

\section{Gravitational waves produced during axion inflation via helical photons.}

In this final section we study a scenario where a pseudoscalar inflaton $\phi$ interacts with a gauge field $F_{\mu\nu}$ through the coupling
\begin{equation}\label{coupl}
{\cal L}_{\phi,F_{\mu\nu}}=-\frac{\phi}{4\,f}\, F_{\mu\nu}\tilde F^{\mu\nu},
\end{equation}
where $f$ is a constant with the dimension of mass. The rolling inflaton excites, through this coupling, quanta of the electromagnetic field, which in their turn source the tensor components of the metric. 

In this section we will first derive the amplitude and the properties of the spectrum of tensor modes generated by this mechanism. We will then study the prospects of direct detection of such modes, focusing on the specific case where $V(\phi)\propto \phi^2$.
\subsection{The amplitude of the tensor modes.}
The production of tensor modes by a pseudoscalar inflaton through gauge field production was discussed in~\cite{Barnaby:2010vf} and, in greater detail, in~\cite{Sorbo:2011rz}, where it was  pointed out that these modes are chiral. We sketch here the derivation of the spectrum of gravitational waves generated by this mechanism, referring the reader to~\cite{Sorbo:2011rz} for a thorough discussion.

In terms of the vector potential ${\bf A}\left(\tau,\,{\bf x}\right)$, defined by $a^2\,{\bf B}=\nabla\times{\bf A}$, $a^2\,{\bf E}=-{\bf A}'$ and neglecting the spatial gradients of $\phi$, the equations for the gauge field subject to the coupling~(\ref{coupl}) read
\begin{eqnarray}
\label{a15}
\left(\frac{\partial^{2}}{\partial \tau^{2}}-\nabla^{2}-\frac{\phi'}{f}\,\nabla\times \right){\bf A}=0,\, \,\,\,\,\,\, \nabla\cdot{\bf A}=0\,,
\end{eqnarray} 
where the prime denotes differentiation with respect to the conformal time $\tau$. We  promote the classical field ${\bf A}(\tau,\,{\bf x})$ to an operator $\hat {\bf A}\left(\tau,\,{\bf x}\right)$, which we decompose on a basis of helicity vectors ${\bf e}_\pm$ as in eq.~(\ref{decoa}).

The functions $A_\pm$ must satisfy the equations $A_{\pm}''+(k^2 \mp k\,\phi'/f)A_{\pm}=0$. Since we are working on an inflating background, we assume the de Sitter metric $a\left(\tau\right)\simeq -1/(H\,\tau)$ and $\dot\phi=\phi'/a=\sqrt{2\,\epsilon}\,H\,M_P\simeq\,$constant. Hence, the equation for $A_\pm$ reads
\begin{equation}
\label{d3}
\frac{d^{2}A_\pm(\tau,\, k)}{d\tau^{2}}+\left(k^{2}\pm 2\,k\,\frac{\xi}{\tau} \right)A_\pm(\tau,\, k)=0\mbox{ ,}
\end{equation} 
where we have defined 
\begin{equation}
\xi\equiv\frac{\dot\phi}{2\,f\,H}=\sqrt{\frac{\epsilon}{2}}\,\frac{M_P}{f}\,\, .
\end{equation}
We will be interested in the case $\xi\ga{\cal {O}}\left(1\right)$, and we assume, without loss of generality, that $\xi>0$. Then $A_-$ stays essentially in vacuum, and we will ignore it from now on. However, the mode function $A_{+}$ develops an instability, and it peaks at momenta $k$ for which $\left(8\,\xi\right)^{-1}\ll |k\,\tau|\ll 2\,\xi$, where it is well approximated by
\begin{equation}
\label{approx1}
A_+(\tau,\, k)\simeq 
\frac{1}{\sqrt{2\,k}}\left(\frac{ k}{2\,\xi\,aH}\right)^{1/4}e^{\pi\,\xi-2\,\sqrt{2\xi \,k/aH}}\,.
\end{equation}
The ``wrong'' sign of the term proportional to $\xi$ in eq.~(\ref{d3}) induces an exponential amplification $\propto e^{\pi\,\xi}$ of the mode function for the photon $A_+$ at sub-horizon scales. Such a large occupation number for the vector field is in its turn a strong source of gravitational waves. Another inflaton-gauge field interaction term that leads to amplification of gauge field modes is the kinetic coupling $f(\phi)\,F_{\mu\nu}\,F^{\mu\nu}$. Such coupling, however, generates only a moderate occupation number of superhorizon photons~\cite{Demozzi:2009fu} and, as a consequence, is not expected to induce an important production of tensors.

We can now study the production of gravitational waves induced by the helical photons. To do so, we plug the expression~(\ref{approx1}) into eq.~(\ref{hijvec}), and we project onto its left- and right-handed components. Of course, one should also take into account the parity-symmetric component of gravitons that is generated by the usual amplification of vacuum fluctuations in de Sitter space. This is uncorrelated from those discussed above so that the overall power spectra of the helicity-$\pm$ components of the graviton can be written for $\xi\gtrsim 2$ (as we will see, this is the regime we are interested in) as
\begin{eqnarray}\label{main}
&&{\cal {P}}^{t,+}=\frac{H^2}{\pi^2\,M_P^2}\,\left(1+8.6\times 10^{-7}\,\frac{H^2}{M_P^2}\,\frac{e^{4\,\pi\,\xi}}{\xi^6}\right)\,\,,\nonumber\\
&&{\cal {P}}^{t,-}=\frac{H^2}{\pi^2\,M_P^2}\,\left(1+1.8\times 10^{-9}\frac{H^2}{M_P^2}\,\frac{e^{4\,\pi\,\xi}}{\xi^6}\right)\,.
\end{eqnarray}
We thus see that,  as a consequence of the violation of parity, the amplitude of the spectra of the left- and the right-handed tensor modes generated by the gauge field differs by a factor $\sim 10^3$.  While the parity violating component could in principle be exponentially large, it was pointed out in~\cite{Barnaby:2010vf} that the gauge field also contributes, through its coupling with the inflaton, to the spectrum of scalar perturbations. This contribution is highly nongaussian, and its amplitude is therefore strongly constrained by the non-observation of nongaussianities in the Cosmic Microwave Background and the Large Scale Structure bispectra. It turns out that the parameter $\xi$, when computed at CMB and at LSS scales, is constrained to be smaller than about $2.6$. This implies that the $\xi$-dependent contribution to the tensor spectra~(\ref{main}), when computed at cosmological scales, is negligibly small. 

Now, the main observation of this section is that the quantity $\xi=\dot\phi/(2\,f\,H)$ is {\em time dependent} and increases as the inflaton rolls down its potential. The condition $\xi\la 2.6$ originates from constraints from CMB and LSS data. Therefore $\xi$ had to be smaller than $2.6$ when LSS scales exited the horizon. However, $\xi$ can be much larger at later times when scales relevant to gravitational wave interferometers left the horizon. Since there are some $40$ efoldings of inflation between the time LSS scales left the horizon and the time LIGO scales left the horizon~\cite{Smith:2005mm}, it is necessary to consider the entire shape of the inflationary potential to know how $\xi$ evolves. In the next subsection we will study in detail, as an example, the parameter space for this scenario in the case of a quadratic inflationary potential. As we will see, there exists a portion of parameter space where Advanced LIGO will be able to observe gravitational waves even if the current bound from nongaussianities is satisfied.
%
\begin{figure}
\centering
    \includegraphics[width=.5\textwidth]{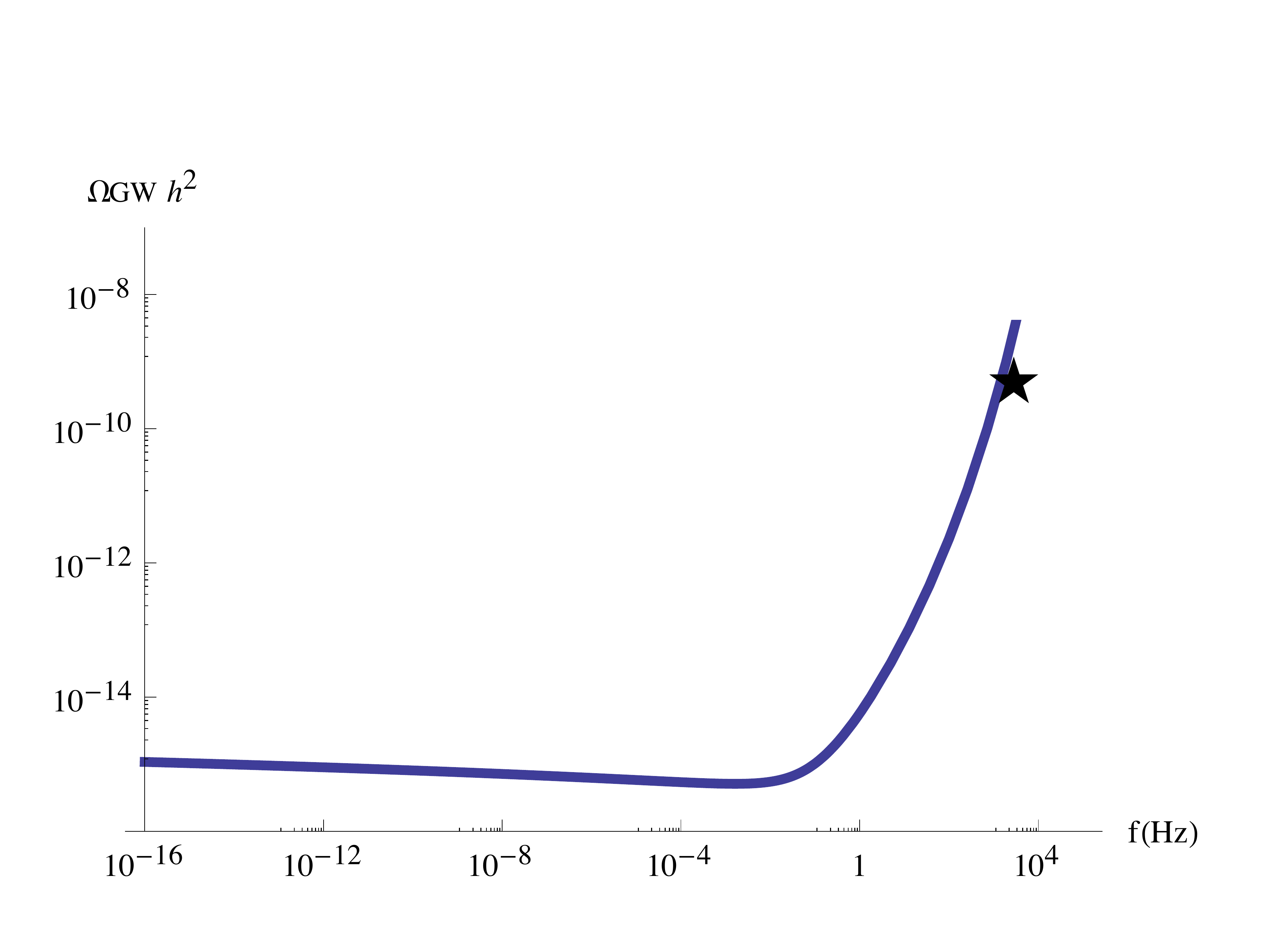}
    \caption{Amplitude of gravitational waves as a function of frequency in the case $\N_c=55$, $\xi_C=2.1$. The star denotes the projected sensitivity of Advanced LIGO.}
\end{figure}
%

\subsection{Gravitational waves from natural chaotic inflation observable by Advanced LIGO.}
In this subsection we study the power spectrum of the gravitational waves~(\ref{main}) in the case where the inflaton potential takes the chaotic form $V(\phi)=\mu^2\,\phi^2/2$. The model of natural chaotic inflation of~\cite{Kaloper:2008fb} (see also~\cite{Kaloper:2011jz}) leads precisely to this situation: a pseudoscalar inflaton with a quadratic potential. Other forms of the potential for a pseudoscalar inflaton were considered in the models of inflation from axion monodromy~\cite{Silverstein:2008sg,McAllister:2008hb,Dubovsky:2011tu}, and we expect that the predictions from these models will not differ significantly from those presented here.

In the case of the chaotic potential $V(\phi)=\mu^2\,\phi^2/2$, the value of $\phi$ during inflation is related to the number $\N$ of efoldings before the end of inflation through $\phi(\N)=2\,\sqrt{\N}\,M_P$. The parameter $\mu$ can be determined by COBE normalization. Denoting by $\N_C$ the number of efoldings corresponding to COBE scales ($47\lesssim \N_C\lesssim 62$ depending on the details of reheating~\cite{Liddle:1993fq,Podolsky:2005bw}), we have
\begin{align}
\mu^2=\frac{6\,\pi^2\,{\cal P}_\zeta}{\N_C^2}\,M_P^2\, ,
\end{align}
where ${\cal P}_\zeta\simeq 2.5\times 10^{-9}$.

Using the slow roll parameter $\epsilon=1/(2\,\N)$, we can write $\xi=\sqrt{\epsilon/2}\,M_P/f$ as $\xi(\N)=\xi_C\,\sqrt{\N_C/\N}$, where $\xi_C$ is the value of $\xi$ computed $\N_C$ efoldings before the end of inflation. We require $\xi_C\la 2.6$ in order not to generate nongaussianities in the CMB which are too large. Note that a comparable constraint on nongaussianities originates from the large scale structures at wavenumbers $k\simeq 0.1$~Mpc${}^{-1}$ (see~\cite{Desjacques:2010jw} for a recent review), which correspond to scales that left the horizon about $\N_C-5$ efoldings before the end of inflation.

Inserting the above expressions into eq.~(\ref{main}), we obtain the following expression of the energy density in gravitational waves as a function of $\N$
\begin{align}\label{omegagw}
\Omega_{GW}\,h^2&=6\times 10^{-14}\,\frac{\N}{\N_C^2}\,\times \\
& \times\left[1+4.2\times 10^{-14}\,\frac{\N^4}{\N_C^5}\,\frac{e^{4\,\pi\,\xi_C\,\sqrt{\N_C/\N}}}{\xi_C^6}\right]\,.\nonumber
\end{align}
 
Since gravitational waves at a frequency $f$ correspond to scales that exited the horizon about $35+\log(f/0.1 {\mathrm Hz})$ efoldings after the COBE scales, we can  plot $\Omega_{GW}\,h^2$ as a function of $f$ for given $\N_C$ and $\xi_C$ by setting $\N=\N_C-35-\log(f/0.1\,{\mathrm Hz})$ in equation~(\ref{omegagw}). We plot in figure 2 the spectrum of gravitational waves for a representative set of parameters.
\subsubsection{Constraints and detectability.}
The parameter space of this system is constrained by the following requirements: first, the backreaction of the electromagnetic modes on the inflating background must be negligible; second, the nongaussianities induced by the same electromagnetic modes through the mechanism discussed in~\cite{Barnaby:2010vf} must be within the  limits imposed both by CMB and by LSS observations.

Backreaction on the inflating background is negligible for~\cite{Anber:2009ua} $e^{2\pi\xi}/\xi^3\ll 700\,V'(\phi)^2/H^6$, i.e.,
\begin{equation}
\left(\frac{\N}{\N_C}\right)^{7/2}\,\frac{e^{2\,\pi\,\xi_C\,\sqrt{\N_C/\N}}}{\xi_C^3}\ll 6\times 10^{10}\,\,.
\end{equation}
This implies that, if we want backreaction to be negligible all the way to a frequency $f$, this condition must be satisfied with $\N=\N_C-35-\log(f/0.1 {\mathrm Hz})$.

As for nongaussianities, the bound of~\cite{Barnaby:2010vf}, when evaluated at COBE scales, gives the constraint $\xi_C< 2.6$.  Comparable bounds also apply to nongaussianities evaluated at Large Scale Structure scales that left the horizon some $5$ efoldings after COBE scales. We therefore impose $\xi(\N_C-5)<2.6$.

We can now discuss the detectability of the tensor modes~(\ref{main}) by gravitational interferometers. To fix ideas we will focus on Advanced LIGO, which will start taking data in the next few years.
%
\begin{figure}
\centering
    \label{fig:sub:a}
   \includegraphics[width=.5\textwidth]{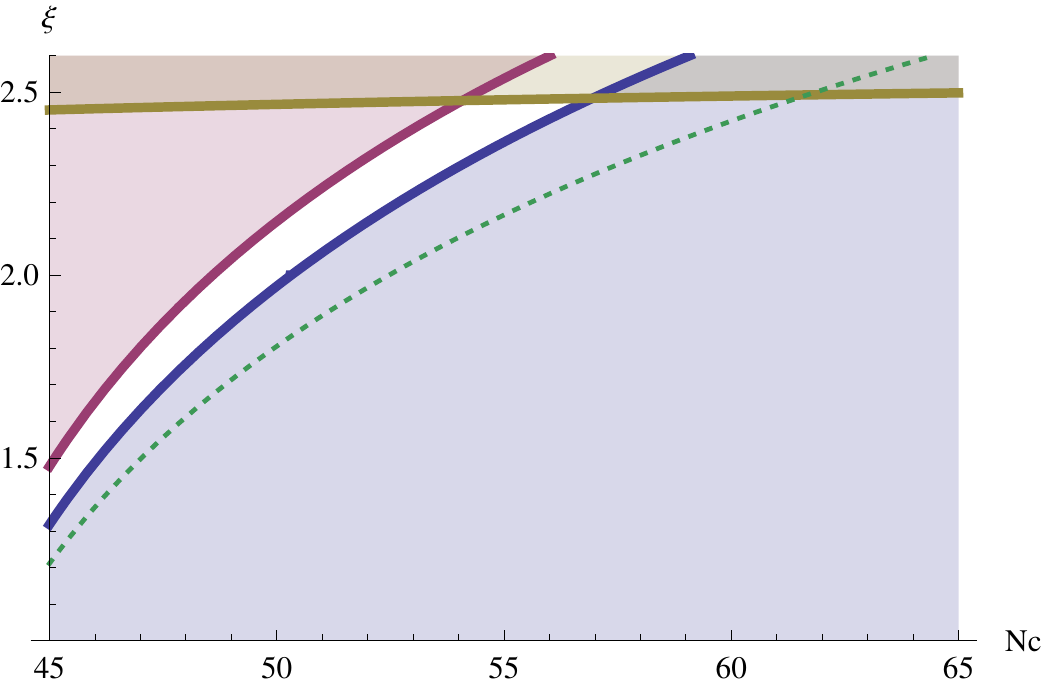}
\caption{Values of $\xi_C$ corresponding to detectable tensor modes by Advanced LIGO, as a function of the total number of efoldings of inflation from the time COBE scales left the horizon. The shaded area on the top left corner corresponds to the region where backreaction cannot be neglected and our analysis cannot be trusted. The shaded area on the top part of the plot corresponds to the region where LSS nongaussianities are too large to be consistent with observations. The shaded area at the bottom corresponds to the region where the amplitude of tensor modes is below the Advanced LIGO detection threshold. Finally, the thinner dotted line corresponds to the lower limit of portion of parameter space accessible to an instrument such as the Einstein Telescope.}
\end{figure}
%
Advanced LIGO is expected to be able to detect $\Omega_{GW}\,h^2=10^{-9}$ at a frequency of about $100$~Hz. The white area in figure 3 corresponds to the region of parameter space where primordial tensor modes might be detected by Advanced LIGO without contradicting the constraints described above. Detection would be possible for values of $\xi_C$ of the order of $2$, corresponding to $f\simeq 10^{17}$~GeV. The shaded area in the upper-left corner of the figure corresponds to a region of the parameter space where  backreaction of the electromagnetic modes on the inflating background cannot be neglected, and an analysis similar to that of~\cite{Anber:2009ua} is needed. While such an analysis is beyond the scope of the present work, it is worth stressing that that region cannot be excluded by existing data and might lead to detectable tensor modes. 

An instrument such as the Einstein  Telescope~\cite{einstein} would be a factor $\sim 10^2$ more sensitive in energy than Advanced LIGO while working at the same frequencies. The thin dotted line in figure 3 delimits the region of parameter space that would lead to a detection of tensors by such an instrument.

Space-based interferometers like LISA, which are sensitive to much lower frequencies, will not be able to detect the tensors~(\ref{main}). Indeed, LISA scales are too close to cosmological scales and the evolution of $\xi$ during inflation is not sufficient to overcome the constraints from nongaussianities. 

We also note that the gravitational waves produced this way will be chiral~\cite{Sorbo:2011rz} and that~\cite{Seto:2008sr} discussed the prospects of  a direct detection of a background of chiral stochastic gravitational waves. These gravitational waves will also have a large~\cite{Adshead:2009bz}, parity violating~\cite{Soda:2011am,Shiraishi:2011st} three point function. Finally, it is nice to speculate how a detection of chiral gravitational waves such as those described by~(\ref{main}) could correlate with the detection of nongaussianities such as those discussed by~\cite{Barnaby:2010vf} at cosmological scales: observation of both the nongaussian signal in the CMB and of gravitational waves at LIGO scales would provide a test of inflation at very different times.

\smallskip

{\em To summarize,} we have shown that particle production during inflation generally leads to a feature in the spectrum of tensor modes. If such a feature happens at the right wavelengths, the tensors might be directly detectable by gravitational interferometers. We have seen that  production of scalar or vector quanta during inflation generates only a modest amount gravitational waves. However, the analysis of section 4 shows that, if the inflaton is an axion coupled to a $U(1)$ gauge field, it can lead to an amplification of gravitational waves that would be directly detectable by Advanced LIGO and/or Advanced Virgo in the next few years.

\section*{Acknowledgments} We thank Neil Barnaby, Ryo Namba and Marco Peloso for valuable comments on the previous versions of the paper (with a special thank to Ryo Namba for finding a mistake in the previous version of section 3 that was significantly affecting the conclusions contained in that section). We also thank Mohamed Anber, Alessandra Buonanno, Chiara Caprini, Jean-Francois Dufaux, David Langlois, Eugene Lim and especially Laura Cadonati for useful discussions.  LS thanks the laboratoire AstroParticule et Cosmologie of the University of Paris 7, for kind hospitality in the course of this work. This work is partially supported by the U.S. National Science Foundation grant PHY-0555304.


\begin{thebibliography}{99}

\bibitem{decigo}
See http://www.amaldi9.org/index.php?option=\\
com\_fabrik\&view=form\&fabrik=15\&usekey=day \&dd=5\&tt=1\&type=flow for a recent presentation

\bibitem{ligo}
http://ligo.org/

\bibitem{geo600}
http://www.geo600.org/

\bibitem{virgo}
https://wwwcascina.virgo.infn.it/advirgo/

\bibitem{tama300}
http://tamago.mtk.nao.ac.jp/

\bibitem{lcgt}
http://gwcenter.icrr.u-tokyo.ac.jp/en/

\bibitem{einstein}
http://www.et-gw.eu/

\bibitem{Smith:2005mm}
  T.~L.~Smith, M.~Kamionkowski, A.~Cooray,
  ``Direct detection of the inflationary gravitational wave background,''
  Phys.\ Rev.\  {\bf D73}, 023504 (2006).
  [astro-ph/0506422].

\bibitem{Chung:1999ve}
  D.~J.~H.~Chung, E.~W.~Kolb, A.~Riotto, I.~I.~Tkachev,
  ``Probing Planckian physics: Resonant production of particles during inflation and features in the primordial power spectrum,''
  Phys.\ Rev.\  {\bf D62}, 043508 (2000).
  [hep-ph/9910437].

\bibitem{Garretson:1992vt}
  W.~D.~Garretson, G.~B.~Field and S.~M.~Carroll,
  ``Primordial magnetic fields from pseudoGoldstone bosons,''
  Phys.\ Rev.\ D {\bf 46}, 5346 (1992)
  [arXiv:hep-ph/9209238].

\bibitem{Adshead:2009bz}
  P.~Adshead, E.~A.~Lim,
  ``3-pt Statistics of Cosmological Stochastic Gravitational Waves,''
  Phys.\ Rev.\  {\bf D82}, 024023 (2010).
  [arXiv:0912.1615 [astro-ph.CO]].

\bibitem{Maldacena:2011nz}
  J.~M.~Maldacena, G.~L.~Pimentel,
  ``On graviton non-Gaussianities during inflation,''
  [arXiv:1104.2846 [hep-th]].
  
\bibitem{Soda:2011am}
  J.~Soda, H.~Kodama and M.~Nozawa,
  ``Parity Violation in Graviton Non-gaussianity,''
  JHEP {\bf 1108}, 067 (2011)
  [arXiv:1106.3228 [hep-th]].
  
\bibitem{Shiraishi:2011st}
  M.~Shiraishi, D.~Nitta, S.~Yokoyama,
  ``Parity Violation of Gravitons in the CMB Bispectrum,''
  [arXiv:1108.0175 [astro-ph.CO]].

\bibitem{Boyle:2007zx}
  L.~A.~Boyle, A.~Buonanno,
  ``Relating gravitational wave constraints from primordial nucleosynthesis, pulsar timing, laser interferometers, and the CMB: Implications for the early Universe,''
  Phys.\ Rev.\  {\bf D78}, 043531 (2008).
  [arXiv:0708.2279 [astro-ph]].

\bibitem{Chialva:2010jt}
  D.~Chialva,
  ``Gravitational waves from first order phase transitions during inflation,''
  Phys.\ Rev.\  {\bf D83}, 023512 (2011).
  [arXiv:1004.2051 [astro-ph.CO]].

\bibitem{Green:2009ds}
  D.~Green, B.~Horn, L.~Senatore, E.~Silverstein,
  ``Trapped Inflation,''
  Phys.\ Rev.\  {\bf D80}, 063533 (2009).
  [arXiv:0902.1006 [hep-th]].

\bibitem{astro-ph/9509049} 
  A.~Berera,
  ``Warm inflation,''
  Phys.\ Rev.\ Lett.\ \ {\bf 75}, 3218  (1995)
  [astro-ph/9509049].

\bibitem{arXiv:0808.1855} 
  A.~Berera, I.~G.~Moss and R.~O.~Ramos,
  ``Warm Inflation and its Microphysical Basis,''
  Rept.\ Prog.\ Phys.\ \ {\bf 72}, 026901  (2009)
  [arXiv:0808.1855 [hep-ph]].

\bibitem{Barnaby:2010vf}
  N.~Barnaby, M.~Peloso,
  ``Large Nongaussianity in Axion Inflation,''
  Phys.\ Rev.\ Lett.\  {\bf 106}, 181301 (2011).
  [arXiv:1011.1500 [hep-ph]].

\bibitem{Barnaby:2011vw}
  N.~Barnaby, R.~Namba, M.~Peloso,
  ``Phenomenology of a Pseudo-Scalar Inflaton: Naturally Large Nongaussianity,''
  JCAP {\bf 1104}, 009 (2011).
  [arXiv:1102.4333 [astro-ph.CO]].

\bibitem{Sorbo:2011rz}
  L.~Sorbo,
  ``Parity violation in the Cosmic Microwave Background from a pseudoscalar inflaton,''
  JCAP {\bf 1106}, 003 (2011).
  [arXiv:1101.1525 [astro-ph.CO]].

\bibitem{Kofman:1997yn}
  L.~Kofman, A.~D.~Linde, A.~A.~Starobinsky,
  ``Towards the theory of reheating after inflation,''
  Phys.\ Rev.\  {\bf D56}, 3258-3295 (1997).
  [hep-ph/9704452].

\bibitem{Dufaux:2007pt}
  J.~F.~Dufaux, A.~Bergman, G.~N.~Felder, L.~Kofman, J.~-P.~Uzan,
  ``Theory and Numerics of Gravitational Waves from Preheating after Inflation,''
  Phys.\ Rev.\  {\bf D76}, 123517 (2007).
  [arXiv:0707.0875 [astro-ph]].

\bibitem{Barnaby:2009mc}
  N.~Barnaby, Z.~Huang, L.~Kofman, D.~Pogosyan,
  ``Cosmological Fluctuations from Infra-Red Cascading During Inflation,''
  Phys.\ Rev.\  {\bf D80}, 043501 (2009).
  [arXiv:0902.0615 [hep-th]].

\bibitem{Finelli:2000sh}
  F.~Finelli, A.~Gruppuso,
  ``Resonant amplification of gauge fields in expanding universe,''
  Phys.\ Lett.\  {\bf B502}, 216-222 (2001).
  [hep-ph/0001231].

\bibitem{Chen:2009zp}
  X.~Chen, Y.~Wang,
  ``Quasi-Single Field Inflation and Non-Gaussianities,''
  JCAP {\bf 1004}, 027 (2010).
  [arXiv:0911.3380 [hep-th]].

\bibitem{Battefeld:2011yj}
  D.~Battefeld, T.~Battefeld, C.~Byrnes, D.~Langlois,
  ``Beauty is Distractive: Particle production during multifield inflation,''
  [arXiv:1106.1891 [astro-ph.CO]].

\bibitem{Demozzi:2009fu} 
  V.~Demozzi, V.~Mukhanov and H.~Rubinstein,
  ``Magnetic fields from inflation?,''
  JCAP {\bf 0908}, 025 (2009)
  [arXiv:0907.1030 [astro-ph.CO]].

\bibitem{Kaloper:2008fb}
  N.~Kaloper and L.~Sorbo,
  ``A Natural Framework for Chaotic Inflation,''
  Phys.\ Rev.\ Lett.\  {\bf 102}, 121301 (2009)
  [arXiv:0811.1989 [hep-th]];
  
\bibitem{Kaloper:2011jz}
  N.~Kaloper, A.~Lawrence, L.~Sorbo,
  ``An Ignoble Approach to Large Field Inflation,''
  JCAP {\bf 1103}, 023 (2011).
  [arXiv:1101.0026 [hep-th]].

\bibitem{Silverstein:2008sg}
  E.~Silverstein, A.~Westphal,
  ``Monodromy in the CMB: Gravity Waves and String Inflation,''
  Phys.\ Rev.\  {\bf D78}, 106003 (2008).
  [arXiv:0803.3085 [hep-th]].

\bibitem{McAllister:2008hb}
  L.~McAllister, E.~Silverstein, A.~Westphal,
  ``Gravity Waves and Linear Inflation from Axion Monodromy,''
  Phys.\ Rev.\  {\bf D82}, 046003 (2010).
  [arXiv:0808.0706 [hep-th]].

\bibitem{Dubovsky:2011tu}
  S.~Dubovsky, A.~Lawrence, M.~M.~Roberts,
  ``Axion monodromy in a model of holographic gluodynamics,''
  [arXiv:1105.3740 [hep-th]].
  
\bibitem{Liddle:1993fq} 
  A.~R.~Liddle and D.~H.~Lyth,
  Phys.\ Rept.\  {\bf 231}, 1 (1993)
  [astro-ph/9303019].
  
\bibitem{Podolsky:2005bw} 
  D.~I.~Podolsky, G.~N.~Felder, L.~Kofman and M.~Peloso,
  Phys.\ Rev.\ D {\bf 73}, 023501 (2006)
  [hep-ph/0507096].

\bibitem{Desjacques:2010jw}
  V.~Desjacques, U.~Seljak,
  ``Primordial non-Gaussianity from the large scale structure,''
  Class.\ Quant.\ Grav.\  {\bf 27}, 124011 (2010).
  [arXiv:1003.5020 [astro-ph.CO]].

\bibitem{Anber:2009ua}
  M.~M.~Anber and L.~Sorbo,
  ``Naturally inflating on steep potentials through electromagnetic
  dissipation,''
  Phys.\ Rev.\  D {\bf 81}, 043534 (2010)
  [arXiv:0908.4089 [hep-th]].

\bibitem{Seto:2008sr}
  N.~Seto, A.~Taruya,
  ``Polarization analysis of gravitational-wave backgrounds from the correlation signals of ground-based interferometers: Measuring a circular-polarization mode,''
  Phys.\ Rev.\  {\bf D77}, 103001 (2008).
  [arXiv:0801.4185 [astro-ph]].


\end{thebibliography}
\end{document}